# In situ investigation of growth modes during plasma–assisted molecular beam epitaxy of (0001) GaN

G. Koblmüller[a)]
*Materials Department, University of California, Santa Barbara, California 93106-5050*

S. Fernandez–Garrido and E. Calleja
*ISOM and Dpto. de Ingeniería Electrónica, Universidad Politécnica, 28040 Madrid, Spain*

J. S. Speck
*Materials Department, University of California, Santa Barbara, California 93106-5050*

Real–time analysis of the growth modes during homoepitaxial (0001) GaN growth by plasma–assisted molecular beam epitaxy was performed using reflection high energy electron diffraction. A growth mode map was established as a function of Ga/N flux ratio and growth temperature, exhibiting distinct transitions between three–dimensional (3D), layer–by–layer and step–flow growth mode. The layer–by–layer to step–flow growth transition under Ga–rich growth was surfactant mediated and related to a Ga adlayer coverage of one monolayer. Under N–rich conditions the transition from 3D to layer–by–layer growth was predominantly thermally activated, facilitating two–dimensional growth at temperatures of thermal decomposition.

[a)]gregor@engineering.ucsb.edu

In recent years, the growth of high precision GaN–based structures for opto– and high power microelectronic applications [1–3] was increasingly accomplished by the powerful growth technique of molecular beam epitaxy (MBE). Essentially, the development of growth surface diagrams [4,5] for the plasma–assisted (PA)MBE growth of (0001) GaN has become highly instrumental for identifying optimum growth regimes to produce device–quality GaN films. Within such growth regimes, the surface properties (i.e. surface roughness and morphology) were similar and given by two important growth parameters, i.e. Ga/N flux ratio and growth temperature.

Low Ga/N flux ratios (Ga/N <1, N–rich growth regime) yielded overall heavily pitted and rough GaN surfaces [5–7]. In contrast, relatively smooth surfaces were acquired under Ga–rich conditions (Ga/N>1), accentuated by a continuous reduction in surface pit density and growth planarization with increasing Ga flux [8,9]. Electron mobilities reached peak values under Ga–rich conditions close to the limit for Ga droplet formation [7], while impurity incorporation was reduced drastically [10].

This enhancement in GaN material properties was interpreted by the existence of a stable Ga surface adlayer [5,10,11] and its strong impact on adatom diffusion [12,13] under Ga–rich conditions. Along its self–surfactant nature, the Ga adlayer was found to form steady–state coverages on the (0001) GaN surface, with values ranging from fractions of one monolayer (ML) to a 2.5–ML–thick bilayer depending on the excess Ga flux present during growth [14].

Despite this progress, the exploration of the surface kinetics within the current GaN growth diagrams has been limited to temperatures below thermal decomposition (i.e. < 750 °C). To further improve the quality of GaN–based devices grown by MBE, it is necessary to extend investigations of the growth conditions and their effects on surface diffusion and crystal growth mode towards much higher temperatures.

In this letter, we establish the correlation between growth parameters and GaN growth mode using two in situ methods, reflection high energy electron diffraction (RHEED) and line–of–sight quadrupole mass spectrometry (QMS). By encompasssing the usually avoided temperature region far beyond 750 °C, we particularly demonstrate the high temperature growth modes and provide a detailed growth mode map for the PAMBE growth of (0001) GaN.

The experiments were carried out in a Gen–II MBE system equipped with standard effusion cells for Ga and a Vecco Unibulb radio frequency plasma source for active nitrogen. As substrate we used a 2–inch (0001)GaN template grown by MOCVD on–axis (with surface vicinality defined < 0.5°) on c–plane sapphire. The substrate temperature was measured by an optical pyrometer. Cross–sectional scanning electron microscopy of thick Ga– and N–limited GaN films grown at low temperatures (680 ºC) was used to calibrate Ga and N fluxes in GaN growth rate units (*nm/min*) [4]. 1nm/min is equivalent to 0.064 ML/s, where 1 ML of GaN corresponds to c/2 = 0.259 nm or $1.14 \times 10^{15}$ GaN/cm$^2$ areal density along the (0001) direction.

The growth mode and surface roughness were analyzed by monitoring the RHEED Bragg spot intensity along the [11$\bar{2}$0] azimuth during homoepitaxial GaN nucleation experiments [15–17]. Simultaneous recording of the postgrowth Ga desorption by quantitative QMS allowed to determine the Ga adlayer coverage formed during the nucleation. The QMS detector was specified with a minimum partial pressure of $1 \times 10^{-13}$ Torr and a time resolution of ~2 sec. To calibrate the desorbing Ga flux in GaN–equivalent growth rate units, the response function of the QMS was measured by exposure of a sapphire wafer at 800 °C to known impinging Ga fluxes (1–20 nm/min) well below the limit for Ga droplet formation [11,14,18]. Note, that all experiments were performed on a single GaN template and were consistently reproducible during successive growths and surface recovery cycles.

Figure 1 shows the QMS–measured desorbing Ga flux (blue datapoints) and the RHEED intensity profile (black curves) during the 50–s long homoepitaxial GaN nucleation for different Ga fluxes at fixed N flux (4.8 nm/min) and temperature (700 °C). Depending on the impinging Ga flux, growth was varied from the N–rich to the Ga–rich growth regime, yielding different QMS desorption profiles and Ga adlayer coverages. According to previous work [14], the maximum Ga desorption during the GaN nucleation was defined as the excess Ga flux desorbing from the surface. Under N–rich conditions, apparently no Ga desorption was measured, since the impinging Ga flux was entirely consumed by the nitrogen atoms (Fig. 1a). For Ga–rich growth (Figs. 1b–d), the desorbing Ga flux increased and matched favorably with the nominally expected excess Ga ($\Phi^{Ga} - \Phi^{N}$). This agreement holds only for moderate excess Ga fluxes, being well below the critical Ga flux for droplet formation at the given temperature (i.e. $\Phi^{Ga} - \Phi^{N} \sim 5$ nm/min at 700 °C) [4].

With increasing Ga flux, the Ga adlayer coverage, as determined by integration of the area below the desorbing Ga flux after each growth pulse (i.e. hatched areas) [14], increased steadily from 0 ML (N–rich growth) to more than a 2 ML–thick Ga bilayer (Ga–rich). Error bars for these coverages are on the order of ±0.2 ML, resulting from the ~2 s time resolution of the QMS. We stress that these values present steady–state Ga coverages, independent of the growth time as far as steady–state growth and desorption was achieved (typically within the first 10–20 s of growth).

Depending on the Ga adlayer coverage, three different RHEED intensity transients upon GaN nucleation were found and associated with three specific growth modes. Under the absence of the Ga adlayer (N–rich growth), the RHEED intensity showed no oscillatory behavior and decreased slightly during growth (Fig. 1a). Concurrently, the RHEED pattern transformed gradually from a streaky pattern (indicative of a smooth GaN template) to a pattern with slight intensity modulations of the Bragg spot towards the end of nucleation (typical for the onset of a roughening surface, see inset). These observations characterize the prevailing 3D growth mode [5,17].

For higher Ga adlayer coverages (0.39 ML and 0.92 ML, respectively), multiple intensity oscillations were observed (Figs. 1b and 1c), with their periodicity consistent with the N–limited GaN growth rate. These characteristics along with persistently streaky RHEED patterns are commonly attributed to a two–dimensional (2D) layer–by–layer growth mode, describing the successive formation of single GaN monolayers [15].

Note, that the RHEED intensity oscillations were quickly damped for Ga adlayer coverages higher than 1 ML, resulting in only one or two lower–frequency oscillations (~3–6s long) (Fig. 1d). Such bi–oscillatory behavior at the onset and end of Ga–rich GaN growth was recently attributed to the layer–by–layer like build–up and desorption of a Ga adlayer (bilayer) interfering with growth such that growth rate oscillations were obscured [19,20]. This transitory behavior at the prevalence of a continuous streaky RHEED pattern was representative for the transition from 2D to one–dimensional (1D) step–flow growth mode [15].

The RHEED intensity transients were further investigated for higher growth temperatures of 750 °C and 780 °C, exhibiting distinct temperature dependencies of the transition Ga fluxes related with the two growth mode boundaries (Fig. 2). In specific, the 3D–to–layer–by–layer growth transition occurred at a boundary Ga flux of $\Phi^{Ga} \sim 3$ nm/min at 750 °C and $\Phi^{Ga} \sim 2$ nm/min at 780 °C, respectively. For the layer–by–layer to step–flow growth transition this boun-dary Ga flux was much higher, i.e. $\Phi^{Ga} \sim 7.5$ nm/min at 750 °C and $\Phi^{Ga} \sim 11$ nm/min at 780 °C, respectively.

Analysis of RHEED transients over a wider temperature range (680 – 780 °C) yielded two Arrhenius plots giving sequences of boundary Ga fluxes for both growth mode transitions. Apparently, with increasing temperature the 3D to layer–by–layer growth transition (Fig. 2c) increased well into the N–rich growth regime (i.e. towards higher excess N fluxes). This demonstrates that at higher temperatures N–rich growth may lead to smooth layer–by–layer growth despite the absence of the Ga adlayer. Indeed, we confirmed sustainable layer–by–layer growth with streaky RHEED patterns up to thicknesses larger than 0.5 μm, especially when growth conditions were selected further away from the transition boundary to 3D growth, i.e. moderately N–rich (0.4 < Ga/N < 1) and higher temperatures (T > 750 °C) [21].

Likewise, the boundary Ga fluxes defining the layer–by–layer to step–flow growth transition increased also with temperature (Fig. 2d). Following the procedure of Fig. 1, we determined for each boundary Ga flux (below and above the growth transition) the absorbed Ga adlayer coverage by QMS, as indicated for each datapoint, and concluded that a steady–state adlayer coverage of approximately 1 ML must be associated with this growth mode transition. Fits were made to the data separating the boundary Ga fluxes for each temperature, which resulted in an apparent activation energy of 1.45 ± 0.2 eV. This suggests that the processes involved with this transition are related to substantial increases in surface diffusion at Ga coverages > 1 ML.

Both growth mode boundaries and their relation to the Ga adlayer coverage are summarized in the growth mode map of Fig. 3. With respect to the three characteristic growth regimes (N–rich, Ga–rich intermediate and Ga–rich droplets) defining the standard PAMBE growth diagram of GaN [4,5], the present map highlights in particular the dependence of growth mode on impinging Ga flux and temperature, including regions of GaN thermal decomposition (> 750 °C) [22]. We note that this map holds only for the given N flux (i.e. 4.8 nm/min) and the slight (<0.5°) surface vicinality of the given (0001) GaN template. Variations in these two parameters are expected to shift the transition boundaries of the growth modes, as they are crucially dependent on deposition rate (i.e. supplied N flux), surface diffusion rate and the nucleation of atomic steps on the surface. Such a map provides therefore not only a guide for the growth of high–quality GaN films but also substantial information about surface diffusion and island formation mechanisms.

Essentially, the current manifestation of the prevalent 3D growth mode under N–rich conditions [4–6] seems to break down for high temperature growth, where thermal decomposition comes into play (> 750 °C). Under these conditions, the presented RHEED intensity oscillations indicated that the thermally activated diffusion seemed fast enough to produce higher adatom mobilities and layer–by–layer growth, as was recently observed even at moderate temperatures below the onset for decomposition [5]. However, in Ref. 5 layer–by–layer growth was unsustainable over time resulting in 3D growth, eventually due to insufficient surface diffusion or limitations of the rather narrow layer–by–layer growth region at these lower temperatures.

Utilizing temperatures of thermal decomposition, we assume that surface diffusion may be further enhanced by the re–evaporating Ga atoms, which adsorb with possibly finite surface lifetimes as Ga adatoms and which migrate readily over the generally N–rich GaN surface [22]. This would yield reduced effective N surface coverages, which in turn may cause a strong decrease in Ga diffusion barrier according to theoretical calculations [12]. Such reduced nitrogen coverage and enhanced surface diffusivity during the competitive growth and decomposition processes have been recently also reported for the PAMBE growth of InN [23], emphasizing the feasibility of high–quality group–III nitride growth by PAMBE under N–rich conditions. However, this scenario holds only for conditions, where the rate of GaN formation is larger than the rate of thermal decomposition. For insufficient Ga fluxes (i.e. < 1 nm/min at 780 °C) GaN formation may be completely suppressed and thermal etching of the GaN surface may result in roughened surfaces.

The second important boundary defining the layer–by–layer to step–flow growth transition, underlies a much stronger influence of surfactant mediated diffusion, given by the critical Ga adlayer of 1 ML. This is in accordance with similar observations of layer–by–layer to step–flow growth mode transitions identified between different Ga–rich growth regimes [5]. All these results agree favorably with the well known autosurfactant effect and the significant reduction in the Ga and N adatom diffusion barriers for Ga adlayer coverages > 1 ML ($E^A$ = 0.4 eV (Ga) and 0.9 eV (N)), as compared to dry GaN surfaces ($E^A$ = 1.8 eV (Ga) and 1.4 eV (N)) or GaN surfaces with a submonolayer coverage of Ga [12,13].

In summary, we demonstrated the correlation between the growth kinetics (Ga/N ratio and temperature) and the three classical homoepitaxial growth modes, 3D islanding, 2D layer–by–layer and 1D step–flow growth mode during the PAMBE growth of (0001) GaN. Utilizing RHEED and line–of–sight QMS, the layer–by–layer to step–flow growth transition was found to be dominated by enhanced surface diffusion rates under Ga adlayer coverages larger than 1 ML. In contrast, the 3D to layer–by–layer growth mode transition was predominantly thermally activated, allowing even 2D growth in the N–rich growth regime at high enough temperatures. We summarized these results in a detailed growth mode map, which allows control of the growth surfaces over a wide range of temperatures, even in regions of thermal decomposition.

This work was supported by DOE SSL project No. DE-FC26-06NT42857 and the Spanish Ministry of Education (MAT2004-2875, NAN04/09109/C04/2, Consolider CSD 2006-19, and the FPU program); the Community of Madrid    (GR/MAT/0042/2004 and S-0505/ESP-0200.

**LIST OF FIGURES:**

Figure 1: QMS–measured desorbing Ga flux and Ga adlayer coverages (blue curves and hatched areas) and RHEED intensity transients (black curves) during 50-s long homoepitaxial (0001) GaN nucleation on MOCVD–grown GaN templates at constant T = 700 °C, N = 4.8 nm/min, but variable Ga fluxes of (a) $\Phi^{Ga}$ = 4 nm/min, (b) $\Phi^{Ga}$ = 5 nm/min, (c) $\Phi^{Ga}$ = 6 nm/min and (d) $\Phi^{Ga}$ = 7 nm/min. RHEED patterns of representative morphologies taken at the end of each growth experiment are shown as insets.

Figure 2: Ga flux dependent RHEED intensity transients during short GaN growth pulses on MOCVD–GaN templates at constant N flux (4.8 nm/min) and temperatures of (a) 750 °C and (b) 780 °C, showing 3D growth (black curves), layer–by–layer growth (blue curves) and step–flow growth (red curves). Arrhenius plots of the boundary fluxes for (c) the 3D to layer–by–layer growth transition under N–rich growth and (d) the layer–by–layer to step–flow growth transition under Ga–rich growth. Note, that a critical Ga adlayer coverage of ~ 1ML determines this transition with an apparent activation energy of ~1.45 ± 0.2 eV.

Figure 3: Summarized map of the growth modes as a function of Ga flux and growth temperature for constant N flux (4.8 nm/min) on on–axis (<0.5° surface vicinality) (0001) GaN, highlighting the interrelation with the three standard GaN growth regimes (Refs. 4,5: N–rich, Ga–rich intermediate and Ga–rich droplets) and Ga adlayer coverages.

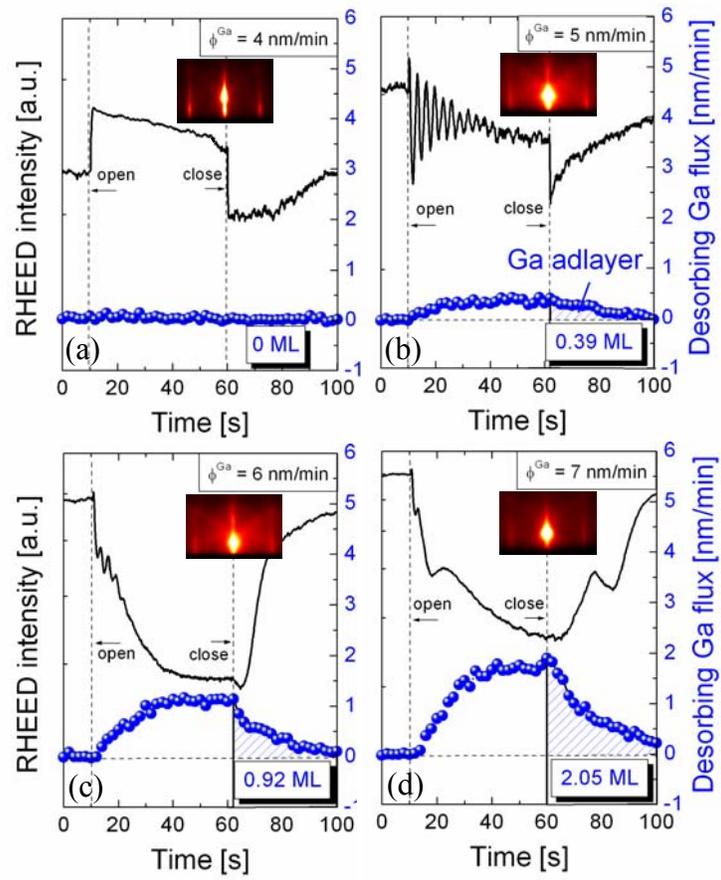

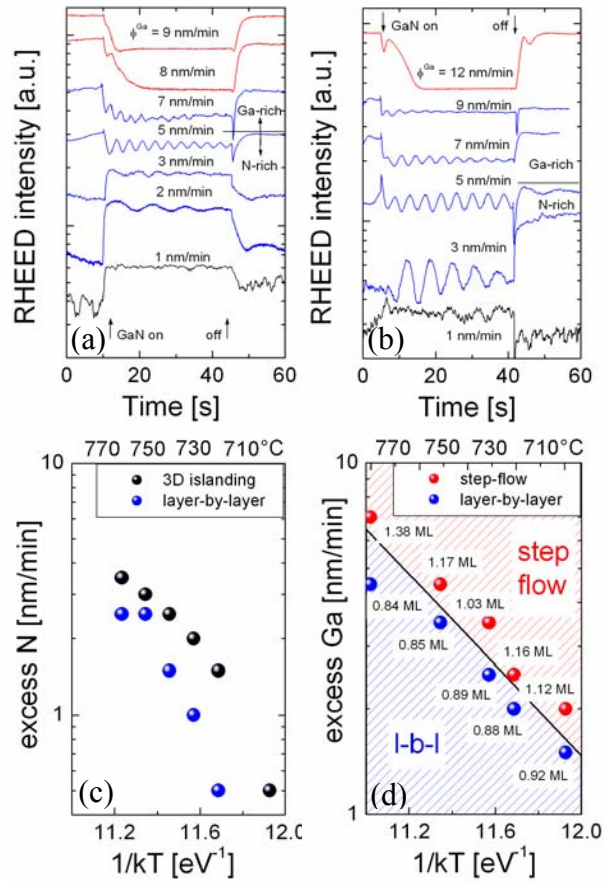

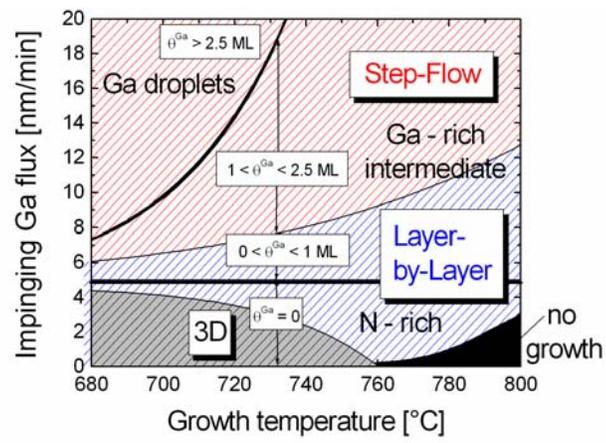

Fig. 3/3
G. Koblmüller

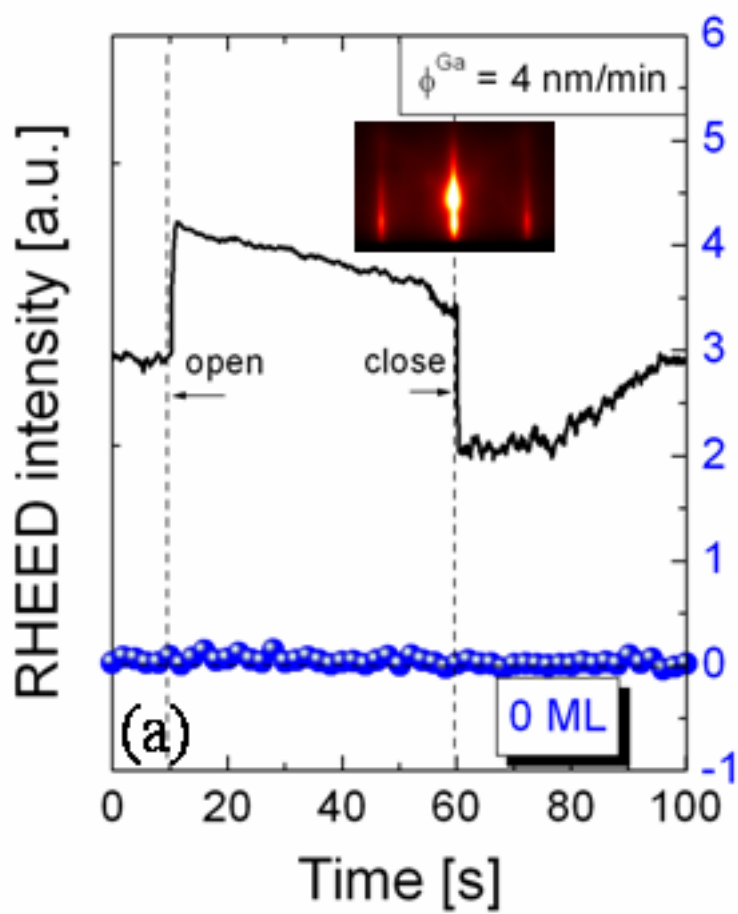

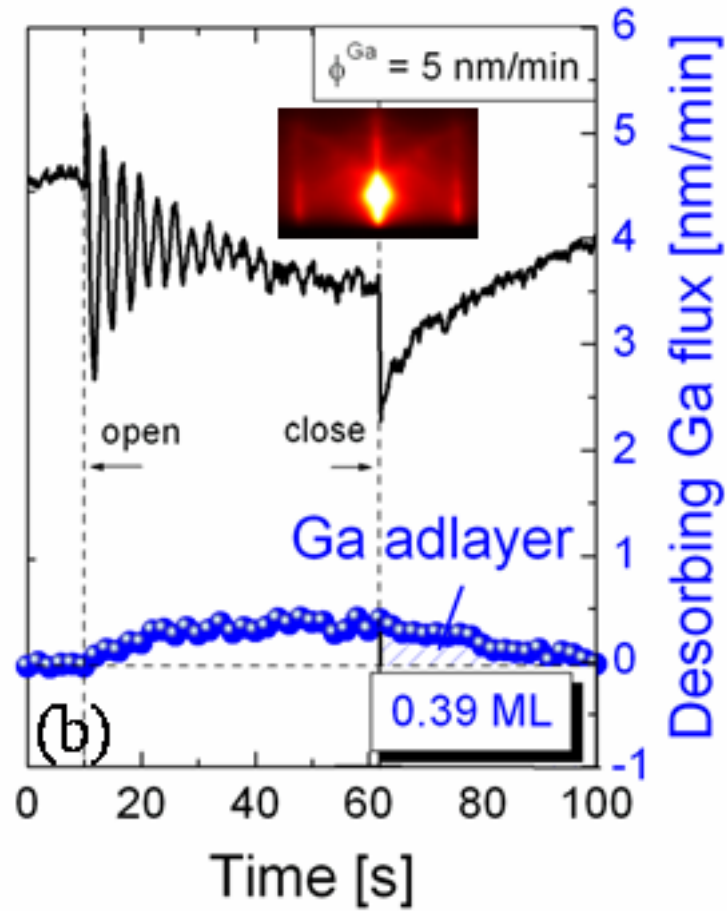

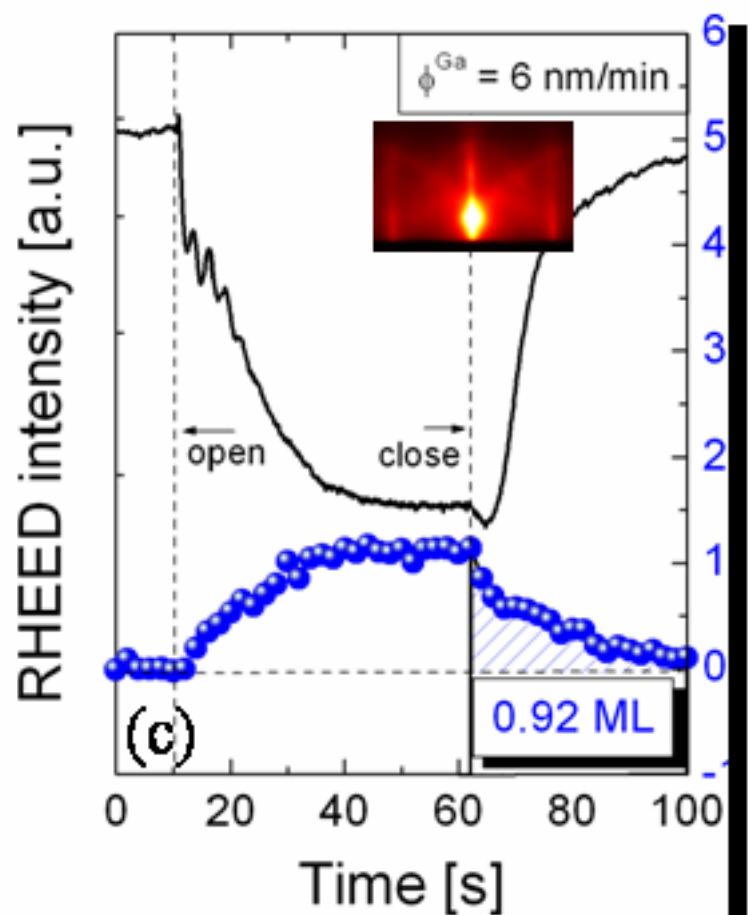

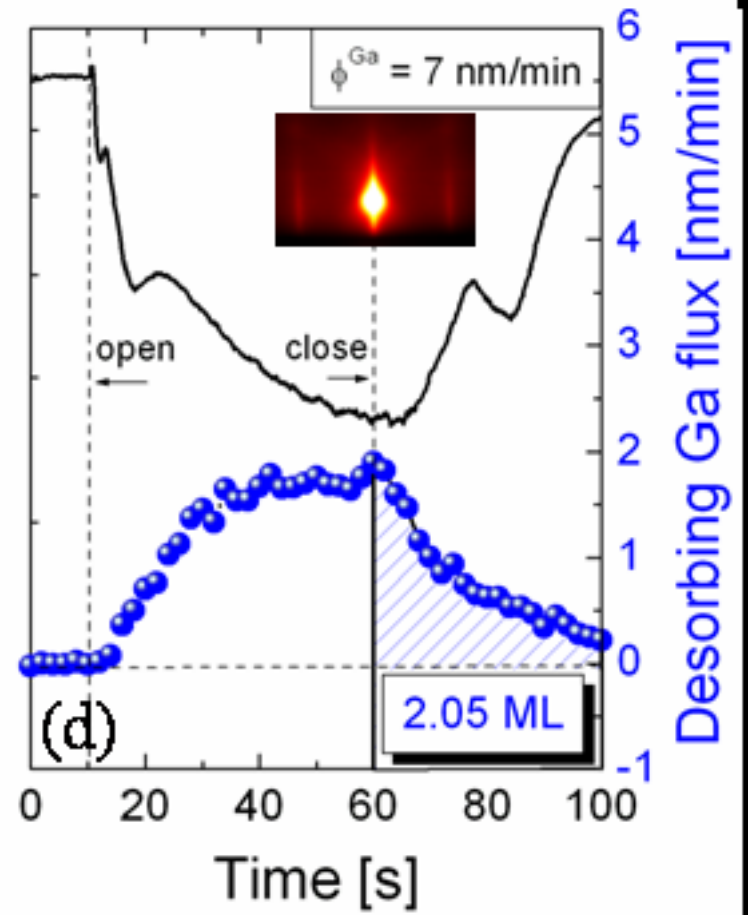

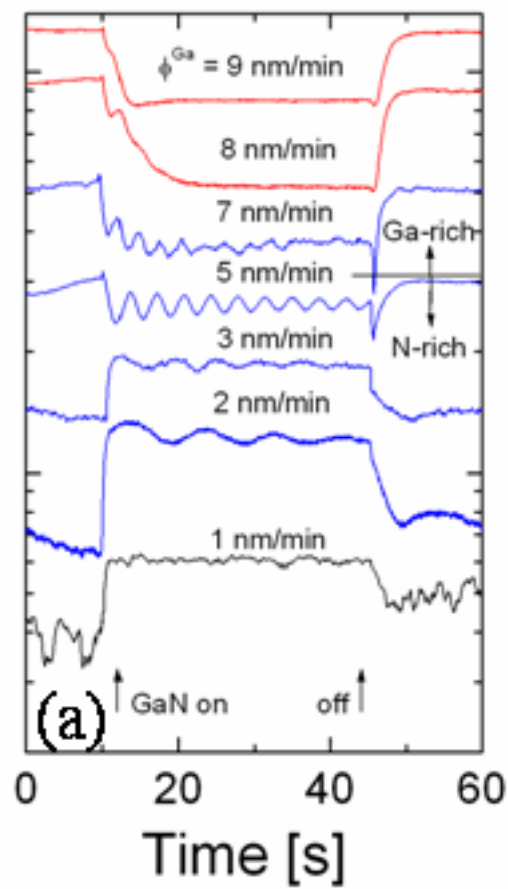

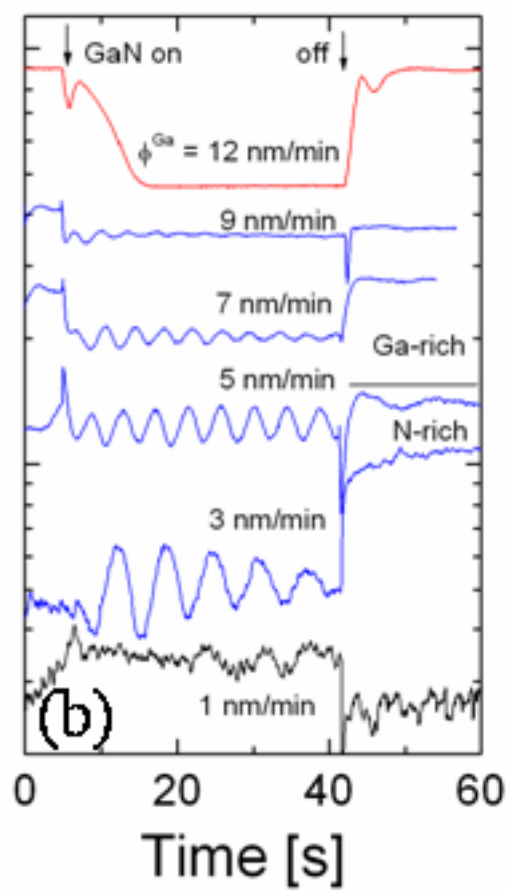

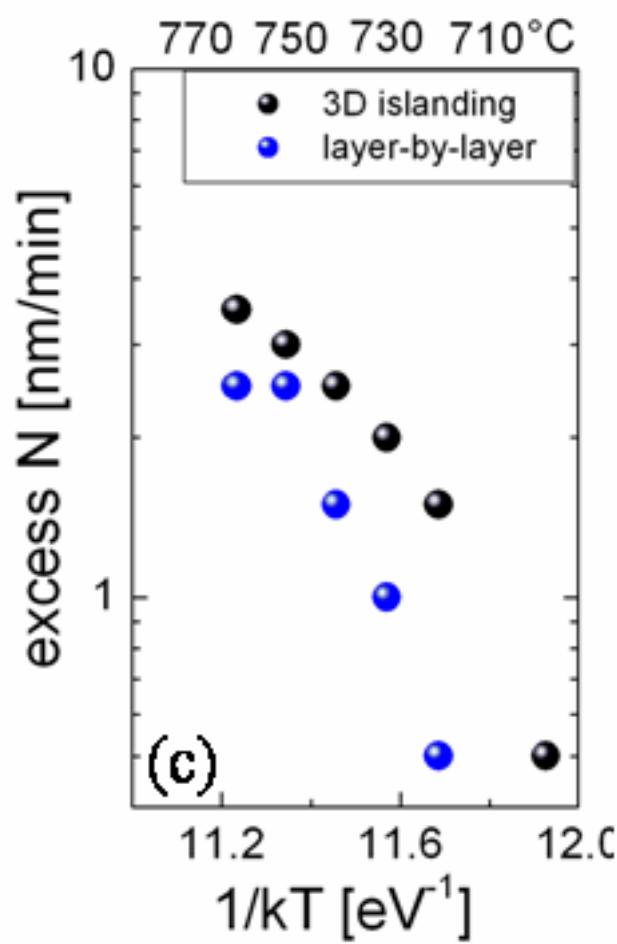

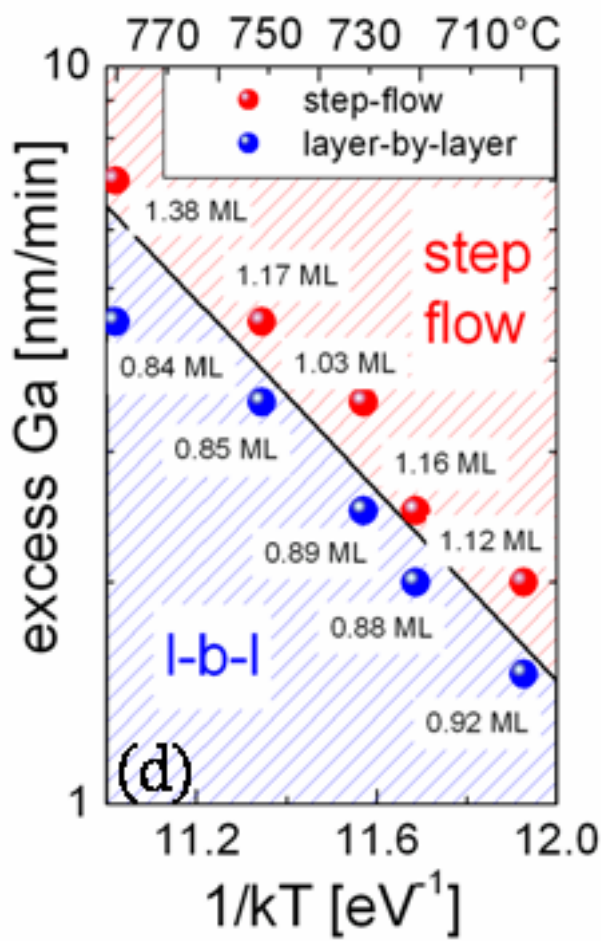

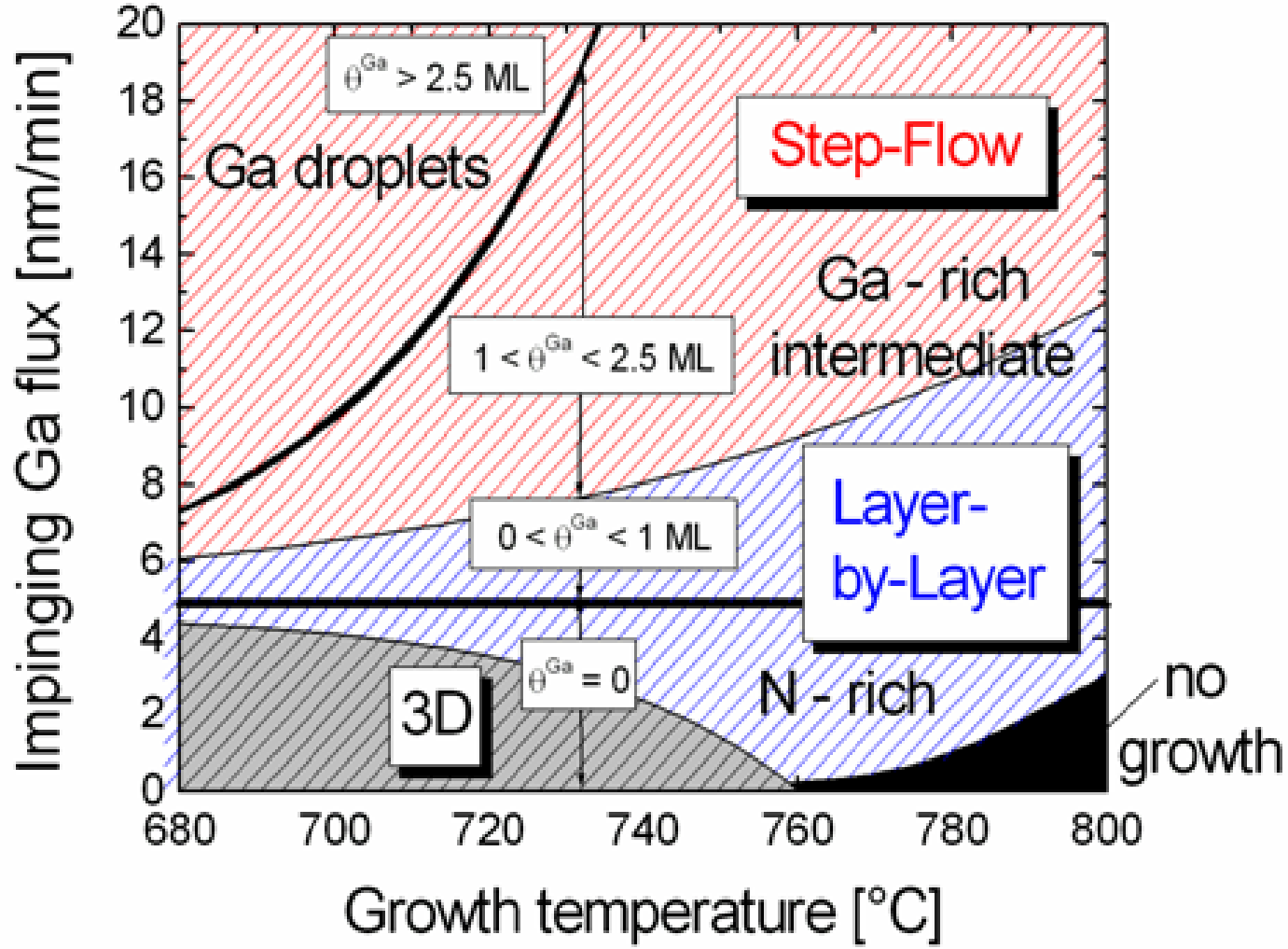